\documentclass[12pt]{iopart}
\newcommand{\BEQ}{\begin{equation}}     
\newcommand{\BEA}{\begin{eqnarray}}
\newcommand{\EEQ}{\end{equation}}       
\newcommand{\EEA}{\end{eqnarray}}

\renewcommand{\vec}[1]{{\bf{#1}}}       

\begin{document}

\input epsf.sty

\title{Cyclic competition of four species: domains and interfaces}

\author{Ahmed Roman, David Konrad, and Michel Pleimling}
\address{Department of Physics, Virginia Tech, Blacksburg, Virginia
24061-0435, USA}
\eads{ \mailto{Michel.Pleimling@vt.edu}}

\begin{abstract}
We study numerically domain growth and interface fluctuations in one- and
two-dimensional lattice systems composed of four species that interact
in a cyclic way. Particle mobility is implemented through exchanges of
particles located on neighboring lattice sites. For the chain we find that 
the details of the domain
growth strongly depend on the mobility, with a higher mobility 
yielding a larger effective domain growth exponent. In two space dimensions, when 
also exchanges between mutually neutral particles are possible, both domain
growth and interface fluctuations display universal regimes
that are independent of the predation and exchange rates.
\end{abstract}

\maketitle

\section{Introduction}
Since the seminal works of Lotka \cite{Lot20} and Volterra \cite{Vol26}, many in-depth studies of
model systems yielded important insights
into food webs and other ecological systems. Using the methods of
non-linear and statistical physics, these investigations allowed to 
make notable progress in our understanding of biodiversity, species
coexistence, and species extinction \cite{Smi74,Hof98,Now06,Sza07,Fre10}. Recent years saw a flurry of studies
of systems composed of multiple species that interact in a cyclic way.
Most of these studies focused on the case of three species
\cite{Fra96a,Fra96b,Pro99,Tse01,Ker02,Kir04,Rei06,Rei07,Rei08,Cla08,Pel08,Rei08a,Ber09,Ven10,Shi10,
And10,Rul10,Wan10,Mob10,He10,Win10,Dob10,He11,Rul11,Wan11,Nah11,Jia11,Pla11,Dem11,He12,Van12,Don12}, 
a special
situation where every species interact with every other species. Only
rather few papers, however, dealt with more realistic cases
where a given species interacts with only
a subgroup of all species living in the same ecological 
environment \cite{Sza07,Fra96a,Fra96b,Dob10,Sil92,Kob97,Fra98,Sza01,Sat02,Sza04,He05,Sza07b,Sza08,
Sor09,Cas10,Dur11,Zia11,Ave12,Dur12,Lut12}.

The case of four species interacting in a cyclic way is characterized by the formation of two alliances
of neutral species, i.e. species that are not in a predator-prey relationship. In the well-mixed situation 
\cite{Cas10,Dur11,Dur12} a variety of orbits in configuration space are obtained in mean field approximation.
Stochastic evolution of finite systems shows a good agreement with the mean field results, but also
reveals interesting additional aspects, most notably those related to various extinction scenarios.
Restricting themselves to predation events, Frachebourg {\it et al.} \cite{Fra96a,Fra96b} studied
one-dimensional coarsening and the related algebraic domain growth. Segregation and the formation
of defensive alliances in two dimensions have been studied in \cite{Kob97,Sat02,Sza04,He05,Sza07b,Sza08}
for a variety of interaction schemes and/or different realizations of particle mobility.

The studies of cyclically competing species have allowed to address a variety of problems,
ranging from biodiversity and species extinction to the formation of space-time pattern. In the present work
we address generic properties of pattern formation during a coarsening process that involves
multiple species. For this we study domain growth and interface fluctuations
in one- and two-dimensional lattice systems when four species compete cyclically. 

Calling the four species $A$, $B$, $C$, and $D$, we prepare fully occupied lattices of $N$ sites
by placing for each of the four species particles
on $N/4$ randomly selected lattice sites. Here $A$ is the predator of $B$, whereas $B$ preys
on $C$, $C$ on $D$, and $D$ on $A$. We update the system by randomly selecting a 
pair of neighboring sites. If the individuals on the selected two sites are in a predator-prey
relationship, then one of the following three events can occur: predation takes place where the prey 
is replaced by a predator, the
two individuals swap places, or nothing happens. This can be written in the following way:
\begin{eqnarray}
A + B \stackrel{k_{AB}}{\to} A + A \nonumber  \\
A+B \stackrel{s_{AB}}{\rightleftharpoons} B+A \nonumber
\end{eqnarray}
and so on. Here, $k_{AB}$ and $s_{AB}$ are the predation and swapping rates between an $A$ and a $B$
particle, with $k_{AB} + s_{AB} \le 1$. If, on the other hand, the two selected individuals are
neutral, as it is the case for $A$ and  $C$ particles, for example, then
they can exchange places with rate $s_{n}$ \cite{Sza07b}.
The results for domain growth and interface fluctuations
discussed in the following have been obtained for cases where all the predation rates
are identical, $k = k_{AB} = k_{BC}= k_{CD} = k_{DA}$. In addition, we also set all the swapping rates
between predator-prey pairs to be equal, $s = s_{AB} = s_{BC}= s_{CD} = s_{DA}$. In this way, we avoid
having a bias in favor of one of the alliances.

The remainder of the paper is organized in the following way. In the next Section we discuss the impact of mobility
on the coarsening process in one dimension. When exchanges between neutral particles are forbidden, an effective algebraic
growth law is encountered where the exponent increases with increasing swapping rate. If we also allow for exchanges between
neutral pairs, the domains grow exponentially fast. The two-dimensional case is the subject of
Section 3. For that case we also find that adding swappings between neutral pairs 
changes qualitatively the domain growth. Indeed, in case of identical predation and exchange rates for all predator-prey pairs,
the system ends up in a steady state where all species coexist.
However, when we allow for exchanges between neutral pairs, coarsening sets in where the domains only contain
mutually neutral partners.
In variance with the one-dimensional case the exponent of the algebraic growth 
in two dimensions is found to be independent of the values $k$ and $s$
of the predation and swapping rates. Section 4 focuses on the interface fluctuations between domains composed of partner-pairs.
We thereby find that the interface fluctuations belong to the Edwards-Wilkinson universality class \cite{Edw82}. 
Finally, in Section 5 we discuss our findings and conclude.

\section{Coarsening in one dimension}

In low dimensions stochastic effects have a remarkable impact on the properties of
non-equilibrium systems, as evidenced for example by the non-trivial nature of many non-equilibrium
phase transitions taking place in one dimension \cite{Hen08}. Coarsening processes in one dimension are usually
substantially different from those taking place at higher dimensions. In a series of papers Frachebourg {\it et al.}
\cite{Fra96a,Fra96b,Fra98} studied the coarsening process in one-dimensional systems composed of multiple species that interact
in a cyclic way. Restricting themselves to immobile particles, they found algebraically growing domains both for three
and four species, but with different values of the growth exponent: 3/4 for three species and 1/3 for four species.
For larger numbers of species, the system rapidly settles into a blocked configuration where neighboring pairs are
non-interacting.

Recent studies of three species systems have shown the importance of particle mobility when cyclic dominance
prevails \cite{Rei07,Pel08,Rei08,Ven10}. Thus in two dimensions mobility induces qualitative
changes in cases where the total number of particles is not conserved. Whereas for low mobility biodiversity
prevails, medium values of the mobility lead to species extinction. Further increasing the mobility finally yields
an enhanced mixing paired with the (re)emergence of species coexistence. In one dimension one finds for the case of
conserved particle number that coarsening takes place at low mobility, paired with a power-law increase
of the average domain size. For high swapping rates, however, complex space-time patterns emerge that ultimately
yield a non-equilibrium steady state where all three species coexist.

\begin{figure}[h]
\centerline{\epsfxsize=1.25in\ \epsfbox{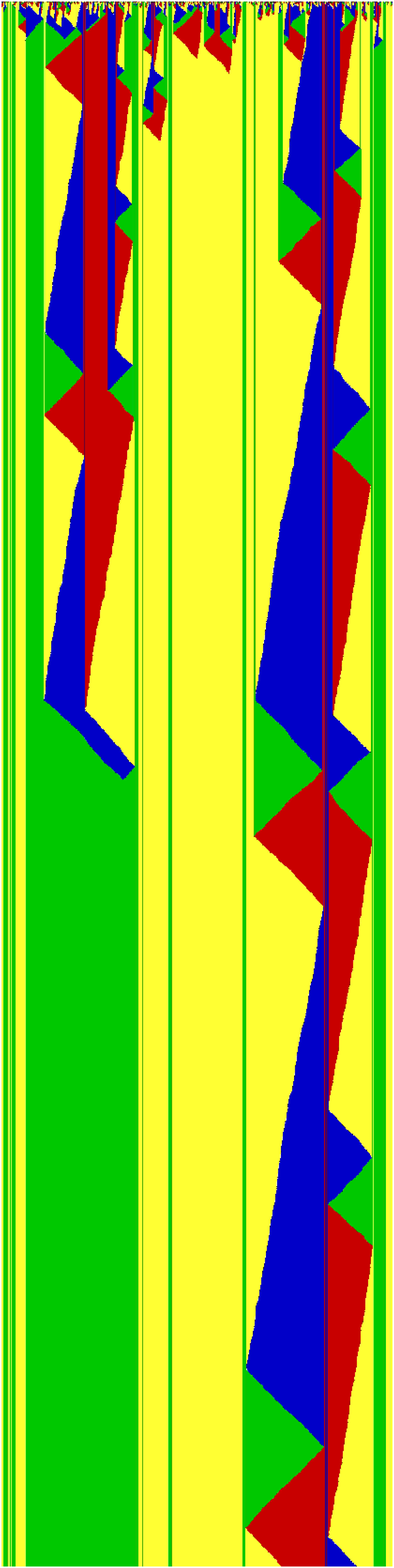} ~~ \epsfxsize=1.25in\ \epsfbox{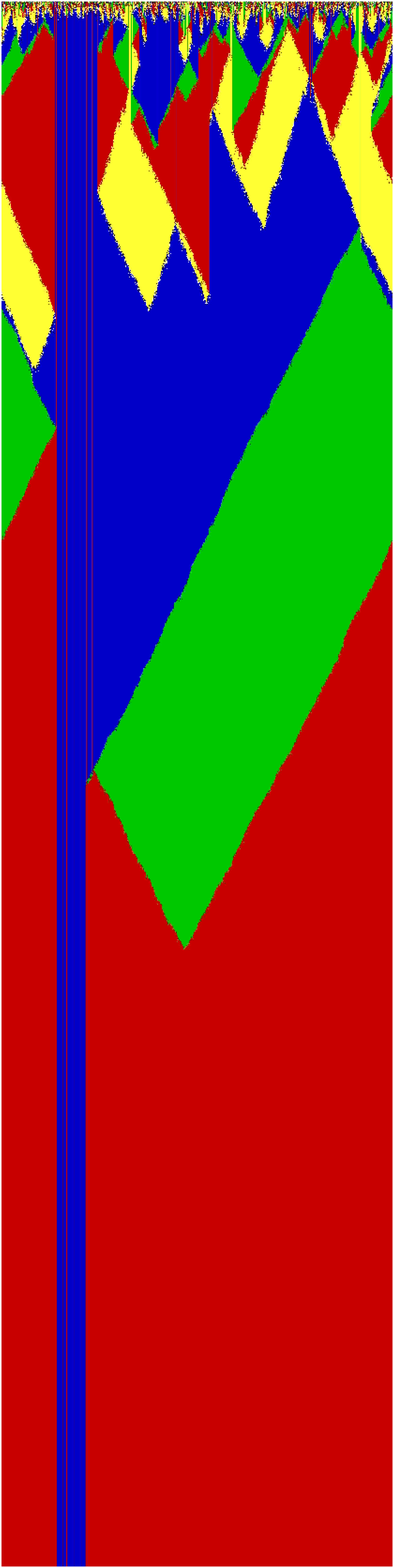}
~~ \epsfxsize=1.25in\ \epsfbox{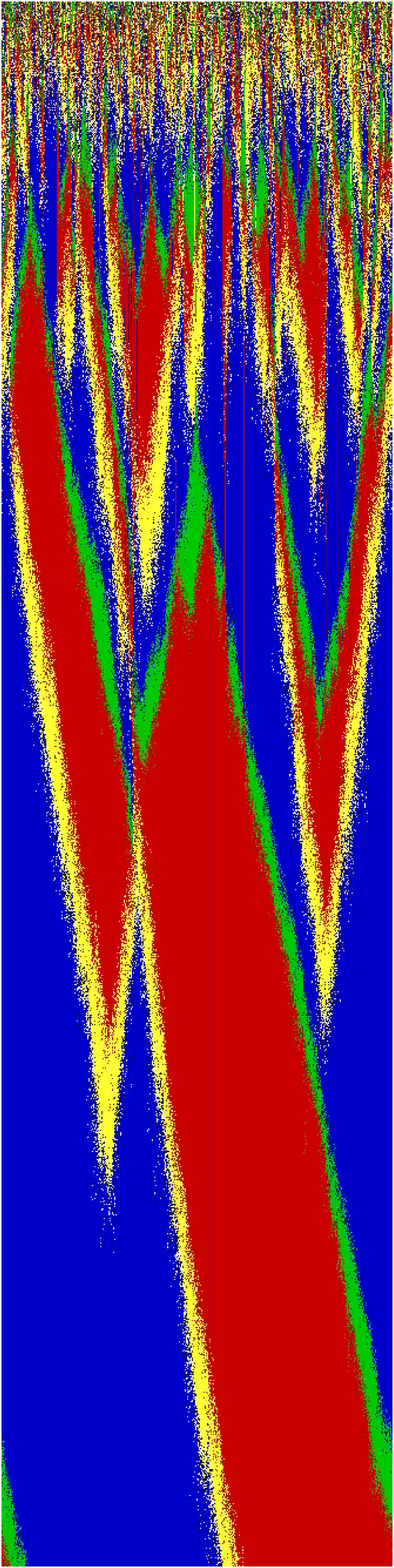}}
\caption{Space-time diagrams (time increases from top to bottom) for one-dimensional systems with three
different predation and exchange rates: $k=0.8$, $s=0.2$, $s_n = 0$ (left), $k=0.1$, $s=0.9$, $s_n =0$ 
(middle), $k=0.01$, $s=0.99$, $s_n = 0$ (right). The system size is $N=2500$, the total simulation time
is $t_{max} = 10000$ MCS. Periodic boundary conditions are used. Blue: $A$ particles, yellow: $B$ particles,
red: $C$ particles, green: $D$ particles.}
\label{fig1}
\end{figure}

In order to get a first impression on how mobility changes one-dimensional coarsening for four species we
can have a look at the space-time diagrams shown in Figure \ref{fig1}. These diagrams have been obtained 
for small systems composed of $N=2500$ sites, with periodic boundary conditions. As usual, time is measured in 
Monte Carlo Steps (MCS), with one MCS corresponding to $N$ proposed updates. In all three cases shown in that
figure, we have set $k + s = 1$. The left diagram shows the typical time evolution of the system for small swapping rates.
As for the case without swapping \cite{Fra96b} we see rapidly the emergence of partner-pair domains (green and yellow
versus red and blue) that are very stable against attacks of the competing alliance. Consider for example the case where
a yellow and a blue domain are in contact. As the $B$ particles (yellow) are the preys of the $A$ particles (blue), the blue domain
expands into the yellow domain until this front meets some $D$ particles (green domains). Now, the roles are changed,
and the blue domain decreases as the $A$ particles are replaced by $D$ particles. This continues until the expanding green
domain encounters a red domain, and so on. This rather complicated coarsening process keeps going on until only one
alliance fills the complete system. Increasing the swapping rate leaves this scenario unchanged (see middle diagram),
but coarsening seems to proceed in an accelerated way. Indeed, a higher swapping rate allows the prey to
survive with a finite probability
the passage through small domains of its predators, so that the protection due to the pairing of neutral 
partners becomes less effective. For very high swapping rate, see the right diagram, it takes some time before 
first domains form.
Once these domains are formed, the coarsening proceeds very rapidly.

\begin{figure}[h]
\centerline{\epsfxsize=5.25in\ \epsfbox{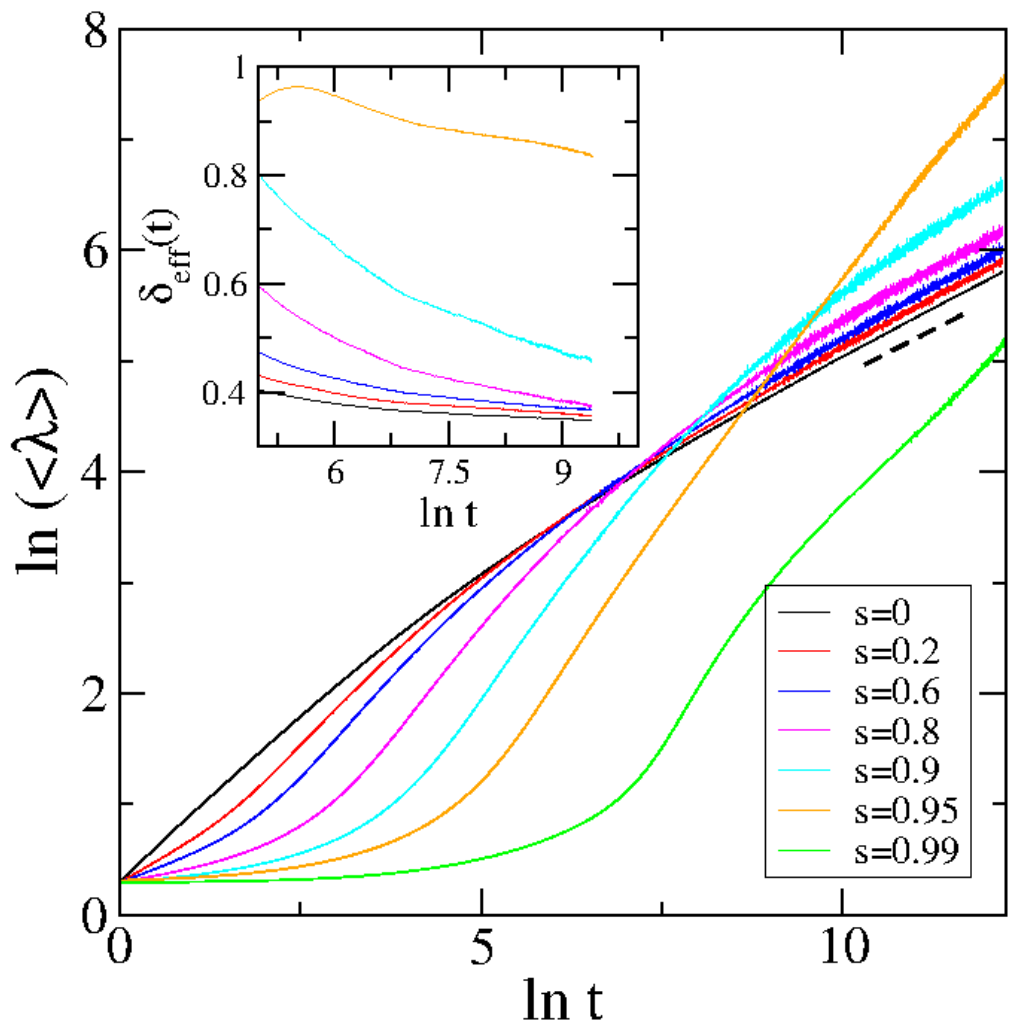}}
\caption{Average domain size as a function of time for different swapping rates $s$, with $k + s =1$.
The systems contain 36000 lattice sites. The dashed line indicates the exponent $1/3$ known exactly
for the case of immobile particles. The inset shows the time-dependence of the effective
exponent (\ref{delta_eff}). The shown data have been obtained after averaging over typically
3000 different realizations of the noise.}
\label{fig2}
\end{figure}

This discussion can be made more quantitative by looking at the average domain size, $\left< \lambda \right>$,
see Figure \ref{fig2}, for various cases with $k + s = 1$. 
We thereby define as average domain size the average length of the segments composed
by only one species. For small swapping rates the domain growth proceeds as for the immobile case,
yielding an algebraic relationship $\left< \lambda \right> \sim t^\delta$, where $\delta = 1/3$ is the exactly
known value when $s=0$ \cite{Fra96b}, see the dashed line. Increasing the value of $s$ has two notable effects.
First, the increase at early times gets slower and slower, thus revealing that the enhanced mobility makes
it more difficult for the initial domains to form. Once ordered segments have formed, however, the growth proceeds
much faster than for the $s=0$ case.
In order to see that we show in the inset the time dependence of the effective exponent
\begin{equation} \label{delta_eff}
\delta_{eff}(t) = \left[ \ln( \langle \lambda \rangle (10 t) ) - \ln (\langle \lambda \rangle (t) ) \right] / \left[ \ln( 10 t) - \ln(t) \right]~.
\end{equation}
We see that this time-dependent effective exponent decreases for increasing
times. In fact, the existing data leave open the possibility that asymptotically the value
1/3 is recovered for extremely long times.

This behavior, which is due to the formation of neutral alliances,
is very different to what is observed in the three species case, see Figures 1 and 2 in \cite{Ven10}. Indeed, as in the three species
case every species interact with every other, one does not have regions formed by domains of mutually  neutral species.
As a result increasing the swapping rate in that case does not yield an 
acceleration of the coarsening, and the domain growth exponent remains unchanged. It is only for very large 
swapping rates that a notable change sets in, yielding a non-equilibrium steady state with constant average domain
size, as the system is then well mixed \cite{Ven10}. We expect the scenario observed in the present paper for the
four species case to be very generic for systems with cyclic dominance composed of multiple species, as long as different
species can form neutral alliances to fight off their predators.

\begin{figure}[h]
\centerline{\epsfxsize=5.25in\ \epsfbox{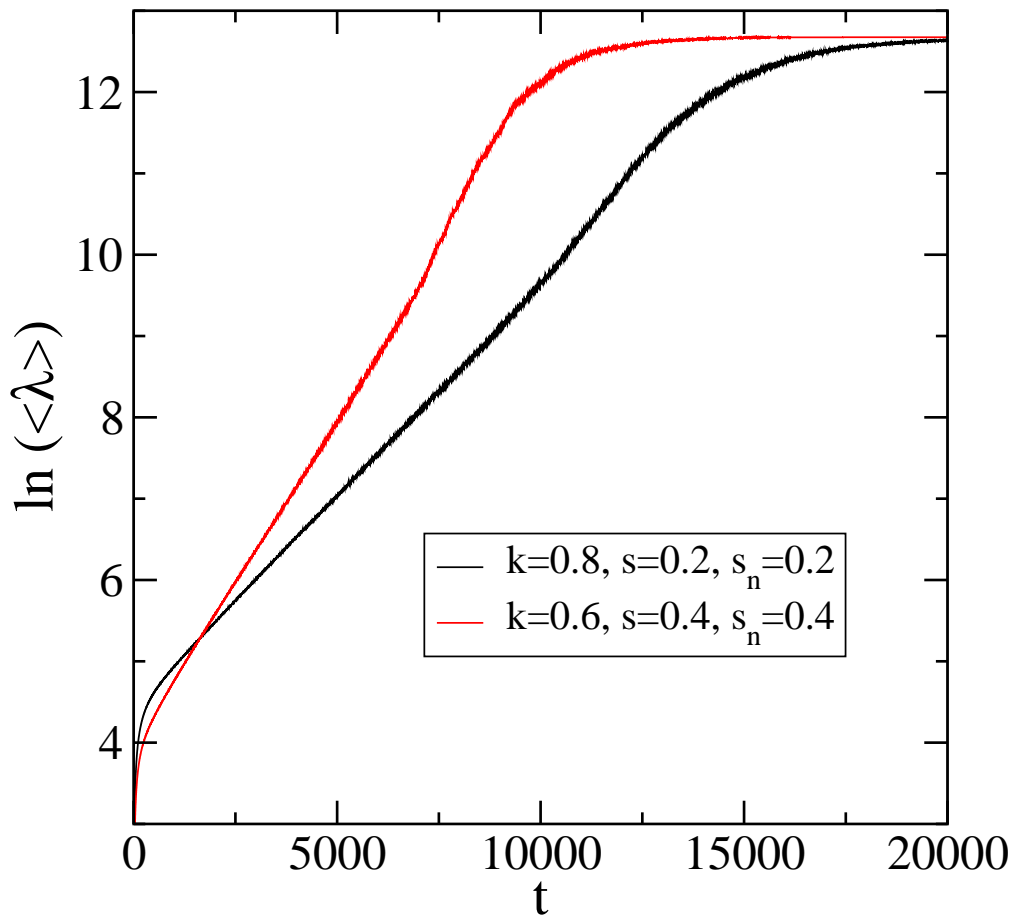}}
\caption{Allowing for swapping between neutral pairs yields an exponential increase of the domain
size, as shown in this linear-log plot. Systems with 36000 sites are considered.}
\label{fig3}
\end{figure}

Following \cite{Sza07b} we can also allow for exchanges between neutral pairs, i.e. swappings
of neighboring $A$ and $C$ particles or $B$ and $D$ particles with rate $s_n$. 
In that case we observe, see Figure \ref{fig3}, that 
the domain growth proceeds in an explosive way, yielding a typical domain size that increases exponentially
with time. Obviously, exchanges between neutral partners weakens the lines of defense
that result from alternating domains formed by the partners, see Figure \ref{fig1},
which drastically increases the probability of dissolving smaller domains. 

\section{Coarsening in two dimensions}

A variety of studies have clarified some of the properties encountered when putting $M > 3$ species
with cyclic dominance on a square lattice. Early on Frachebourg and Krapivsky \cite{Fra98}
found that in the absence of particle mobility systems with more than 14 species end
up in a frozen state where all particles are surrounded by neutral partners. For the case we are 
interested in, namely that of four species, the system settles into a steady state where all four species
coexist. Allowing for mobile particles through the diffusion via unoccupied sites yields a qualitative
change \cite{Sza04}. For small densities of empty sites the behavior is similar to that of immobile particles, and
all four species coexist. For higher densities, however, a coarsening regime is revealed where 
partner-pairs are competing against each other, yielding the formation of domains that contain only
neutral species. A similar transition between a regime of species coexistence and a regime where species extinction
proceeds through a coarsening process is observed for a completely filled lattice when 
exchanges between neutral partners take place \cite{Sza07b}. No exchanges between predators and preys
were allowed in that work.

In the following we study the coarsening process taking place in our system when
exchanges between neutral partners are allowed in addition to interactions (predation and exchange)
between predators and preys. We thereby focus on two different sets of
rates, namely (1) $k=0.8$ and $s = s_n = 0.2$ and (2) $k = 0.2$ and $s = s_n = 0.8$, similar
results being obtained for other values of predation and swapping rates. 
We stress that, in agreement with \cite{Sza07b}, coarsening does not take place in absence
of exchanges between neutral partners, but instead the system then ends up in a stationary state
with coexitence.

\begin{figure}[h]
\centerline{\epsfxsize=2.25in\ \epsfbox{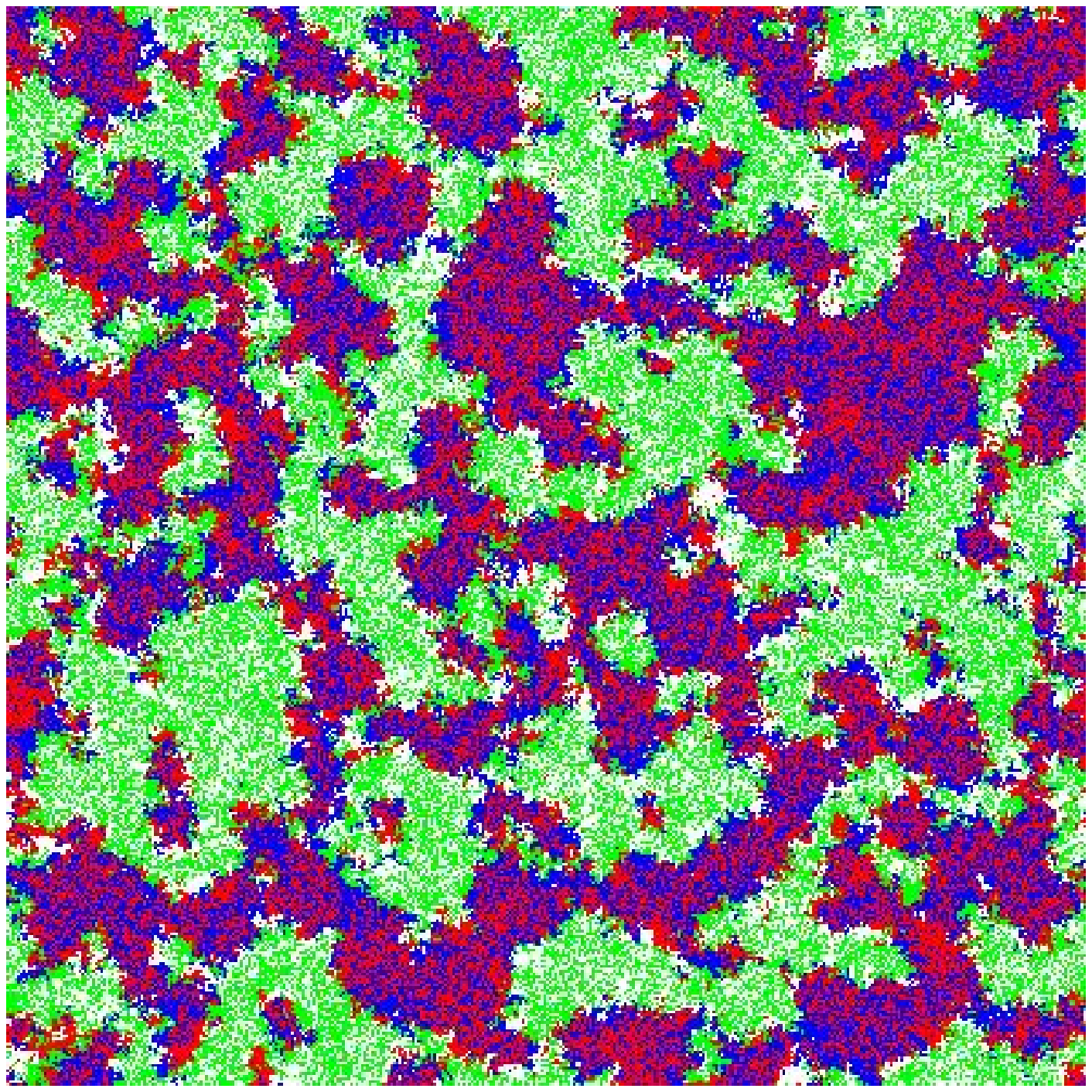} ~~ \epsfxsize=2.25in\ \epsfbox{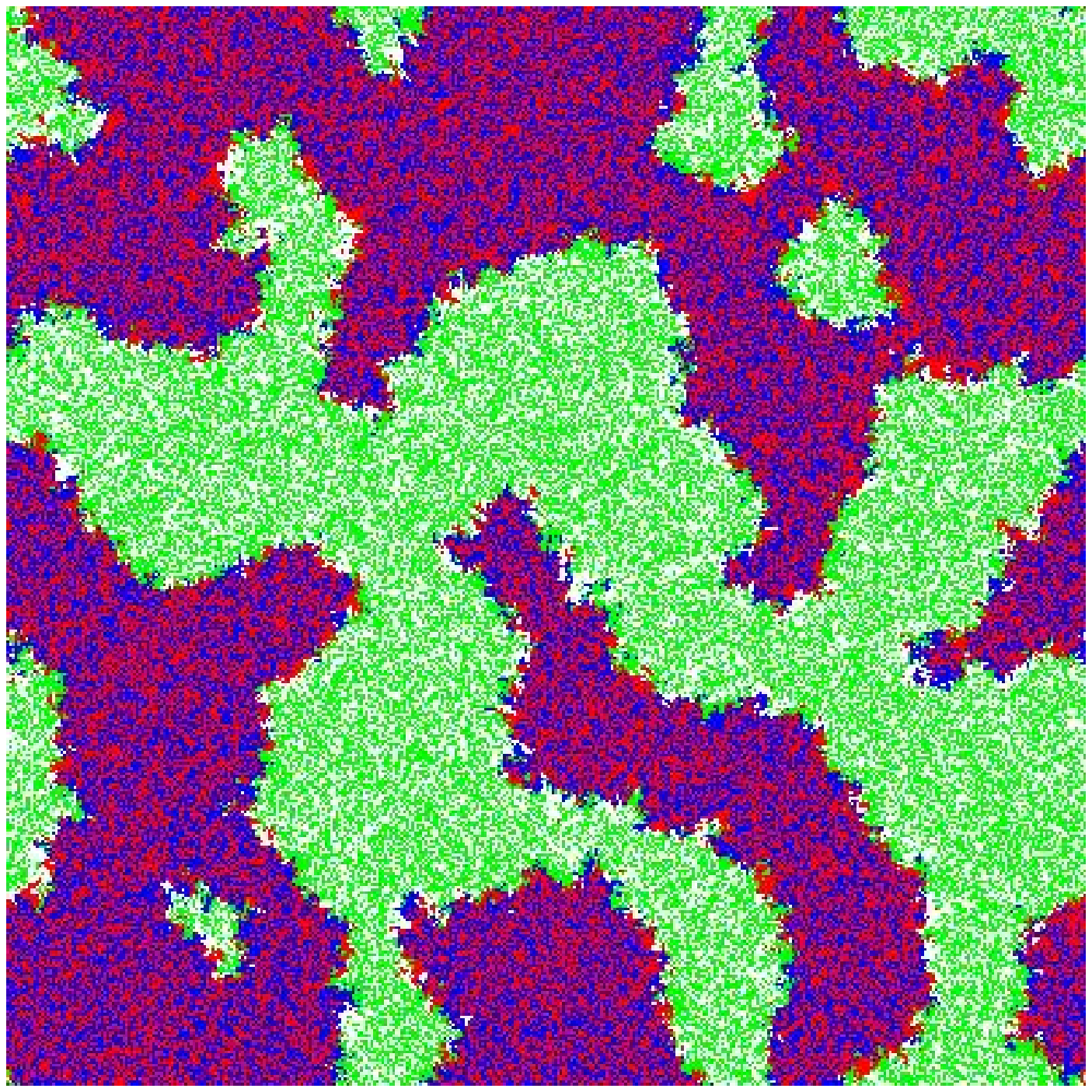}}
\vspace*{0.6cm}
\centerline{\epsfxsize=2.25in\ \epsfbox{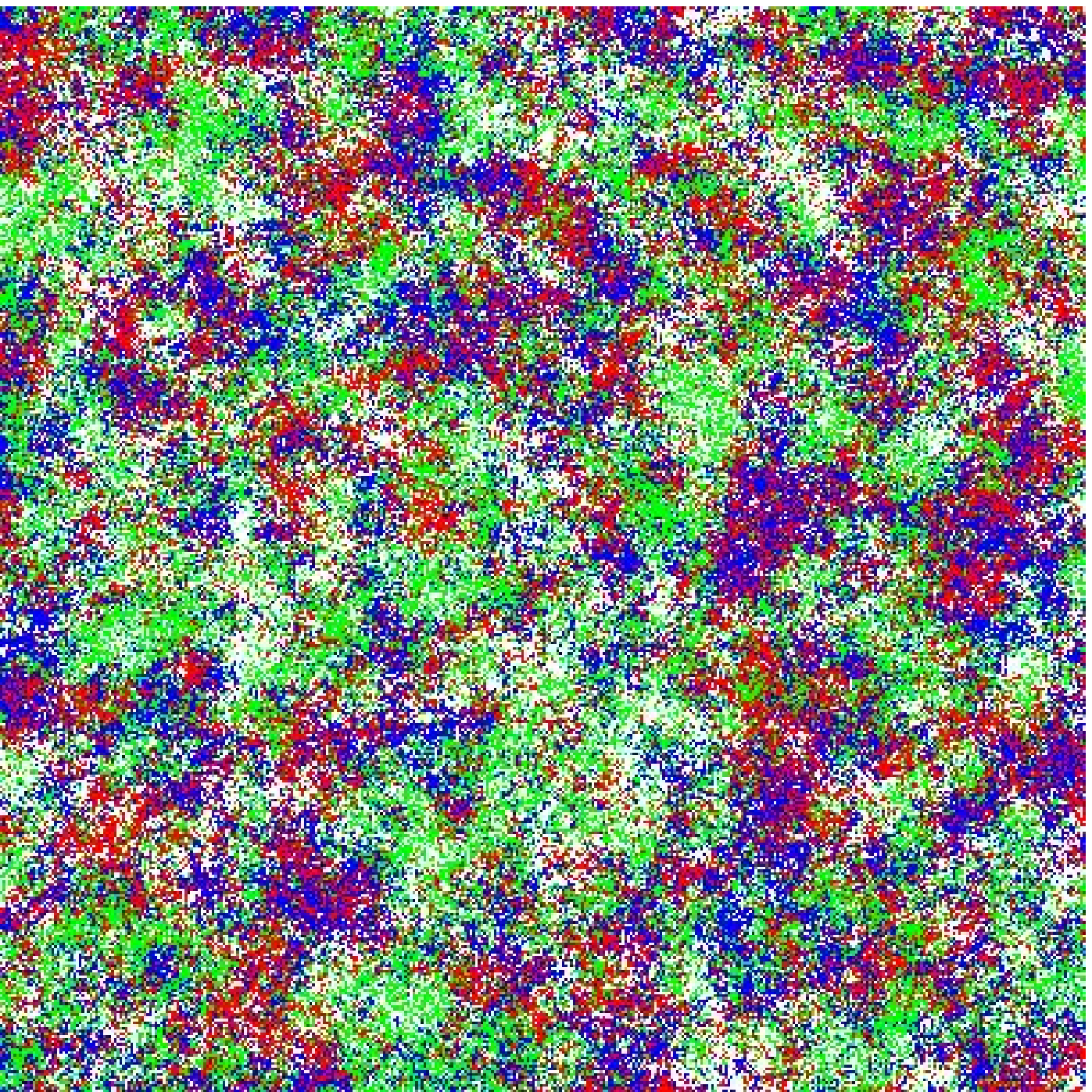} ~~ \epsfxsize=2.25in\ \epsfbox{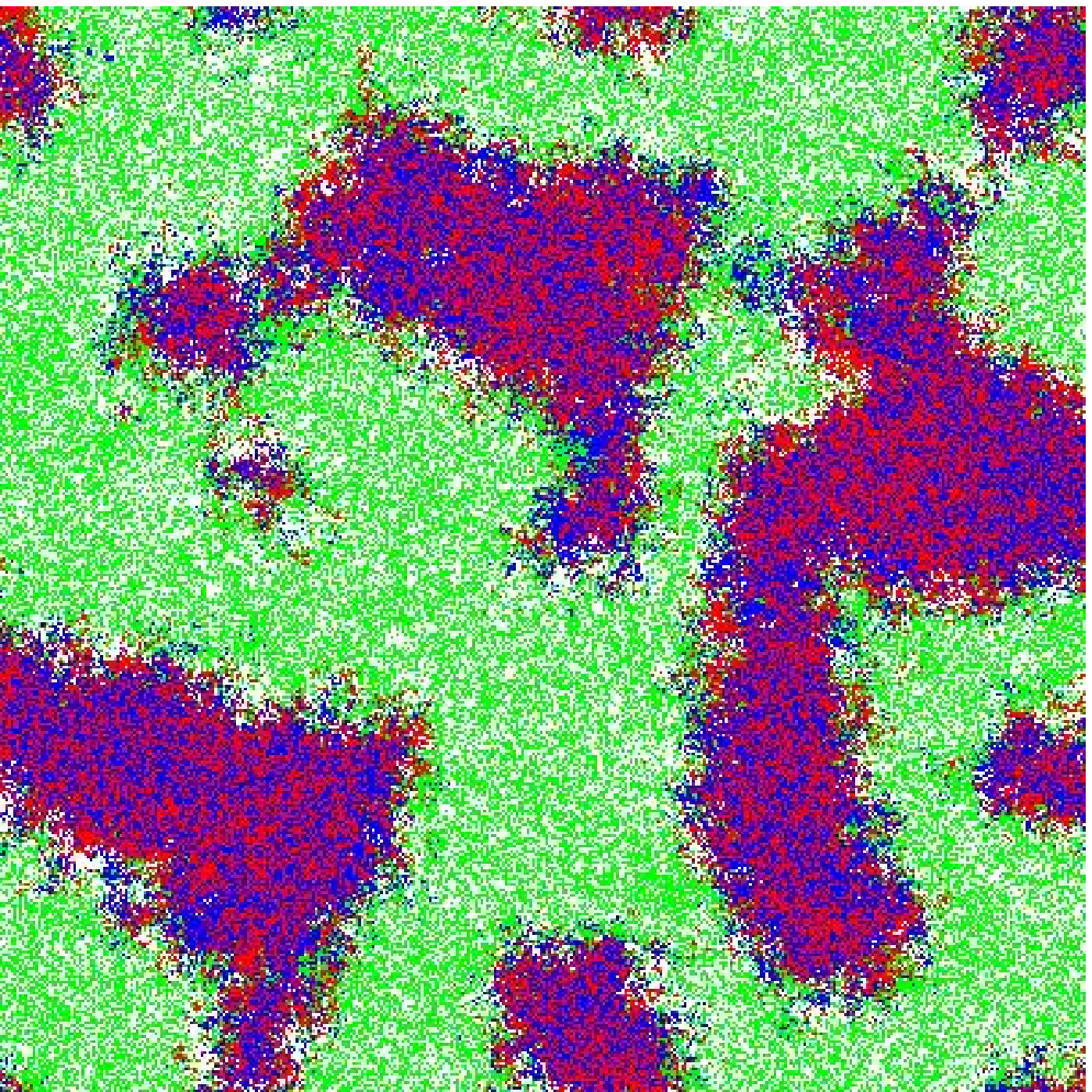}}
\caption{Configurations during coarsening in two dimensions. Top row: $k=0.8$ and $s = s_n = 0.2$, bottom
row: $k = 0.2$ and $s = s_n = 0.8$. The left column shows the systems 100 MCS after initialization,
the snapshots in the right column are taken after 500 MCS. The system size is $400 \times 400$, and 
periodic boundary conditions are used.}
\label{fig4}
\end{figure}

Let us again start by having a look at some configurations, see Figure \ref{fig4}, where the two different
cases are displayed. The first row shows two snapshots, one after 100 MCS and one after 500 MCS, for 
$k=0.8$ and $s = s_n = 0.2$. As expected, one observes the formation of domains that are composed
of mutually neutral species, namely $A$ and $C$ (blue and red) or $B$ and $D$ (white and green).
These domains then coarsen until, eventually, only one alliance will prevail in our finite system.
For the case $k = 0.2$ and $s = s_n = 0.8$, shown in the second row, domains are not yet well formed
after 100 MCS. At a later time we again have coarsening, where smaller domains give place to larger
domains, but the boundaries of the domains are very fuzzy. These effects are of course due
to the enhanced swapping rate and are similar to those observed in the one-dimensional case,
see Figure \ref{fig1}.

\begin{figure}[h]
\centerline{\epsfxsize=5.25in\ \epsfbox{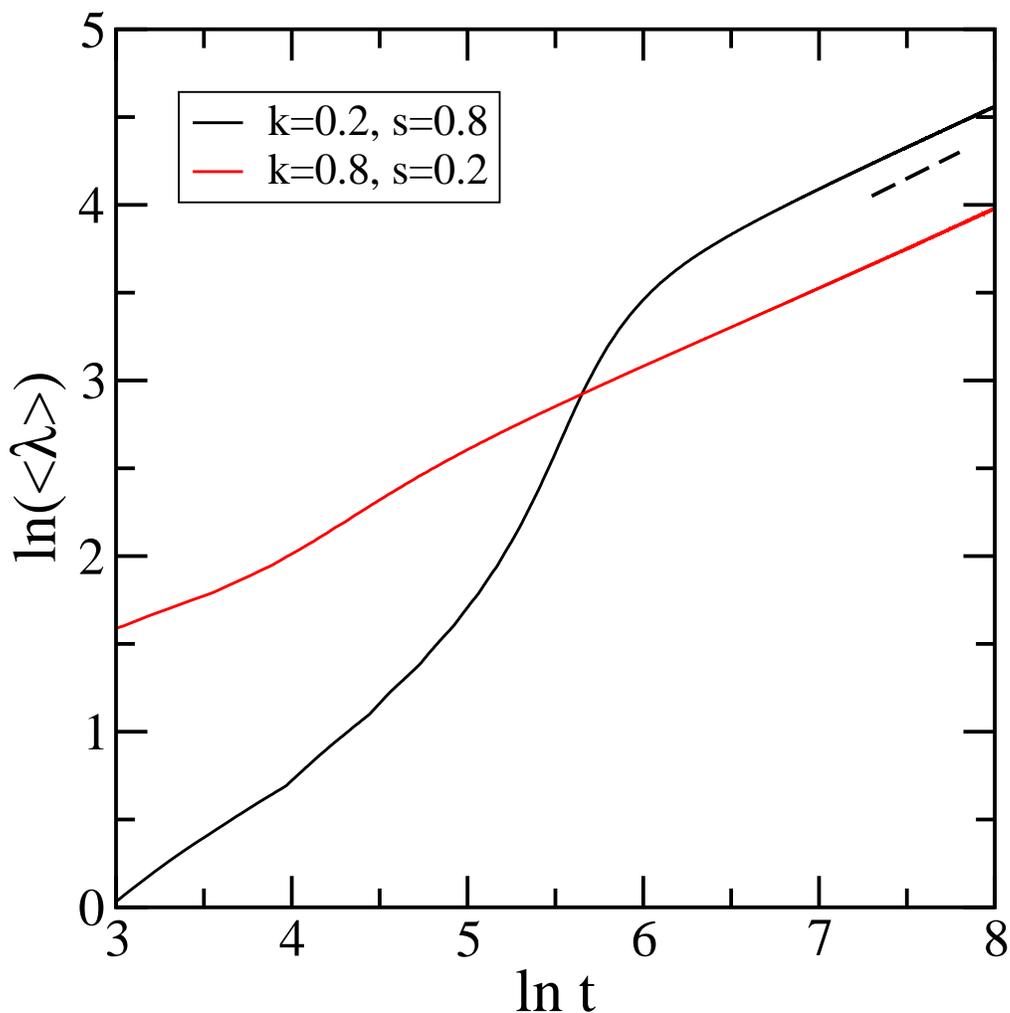}}
\caption{Average correlation length as a function of time for the same two sets of
predation and exchange rates as those used in Figure \ref{fig4}. The dashed line
indicates the slope 1/2. The system size is 
$800 \times 800$, and the data result from an average over 8000 independent runs.}
\label{fig5}
\end{figure}
 
Due to the fact that the boundaries of the domains are not very sharp, we determine
the typical length in our system through a study of the correlation function,
\begin{equation}
C(t, \vec{r}) = \sum\limits_i \left[ \left< n_i(t,\vec{r}) n_i(0,\vec{0}) \right> - 
\left< n_i(t,\vec{r}) \right> \left< n_i(0,\vec{0})
\right> \right]~.
\end{equation}
Here $n_i(t,\vec{r})$ is an occupation number that is 1 when at time $t$ site $\vec{r}$ is occupied
by a particle of species $i$ and 0 otherwise. $\left< \cdots \right>$ indicates an average over
initial states and over the realizations of the noise.
We then define the correlation length $\lambda(t)$ as the distance at which $C(t,\vec{r})$ drops to one third  of the
value it has at $\vec{r} = \vec{0}$. The result of this analysis is shown in Figure \ref{fig5}.
At short times the correlation length is much smaller for the larger swapping rate, see the black line
in the figure, in accordance
with our observation that in that case it takes longer for domains to form. 
Once these domains are formed, the coarsening undergoes a regime of rapid growth before entering
the asymptotic regime. This behavior is in fact very similar to what happens in the $d=1$ case,
see Figure \ref{fig2}. However, whereas in one dimension the asymptotic regime is characterized by an algebraic
growth with mobility dependent exponents, in two dimensions the asymptotic growth is the same for the studied
cases, with $\left< \lambda(t) \right> \sim t^{1/2}$. This universal exponent 1/2, which has also been observed
in other variants of the four species model on the square lattice \cite{Sza07}, is expected from the
Allen-Cahn law for curvature-driven coarsening \cite{Gre88,Lau88,Bra94}. 

\section{Interface fluctuations}

Even so the same growth law prevails in two dimensions for different reaction rates, the boundaries
between the different domains can look very differently, see the configurations shown in
Figure \ref{fig4}. Whereas for small swapping rates these boundaries are very sharp,
they get increasingly fuzzier when increasing this exchange rate. It is therefore {\it a priori} not clear that  
interfaces for different rates have the same properties.

\begin{figure}[h]
\centerline{\epsfxsize=2.25in\ \epsfbox{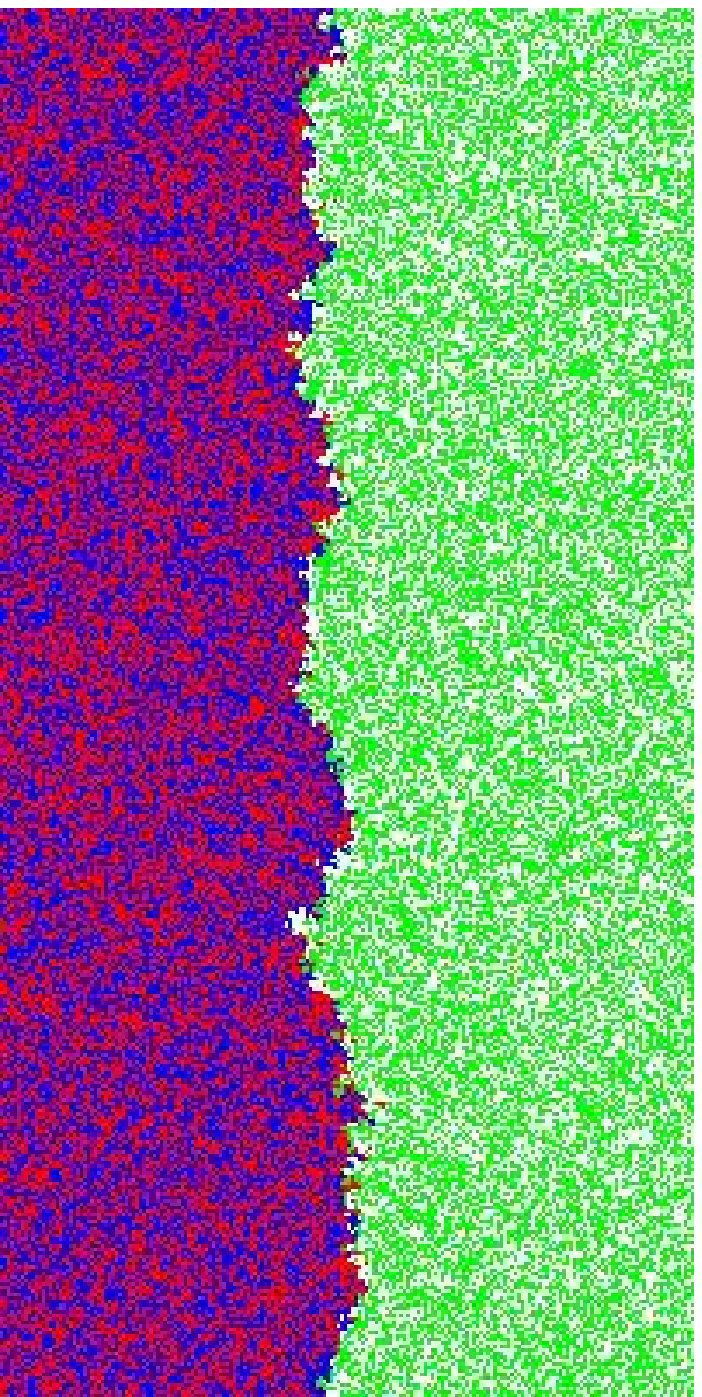} ~~ \epsfxsize=2.25in\ \epsfbox{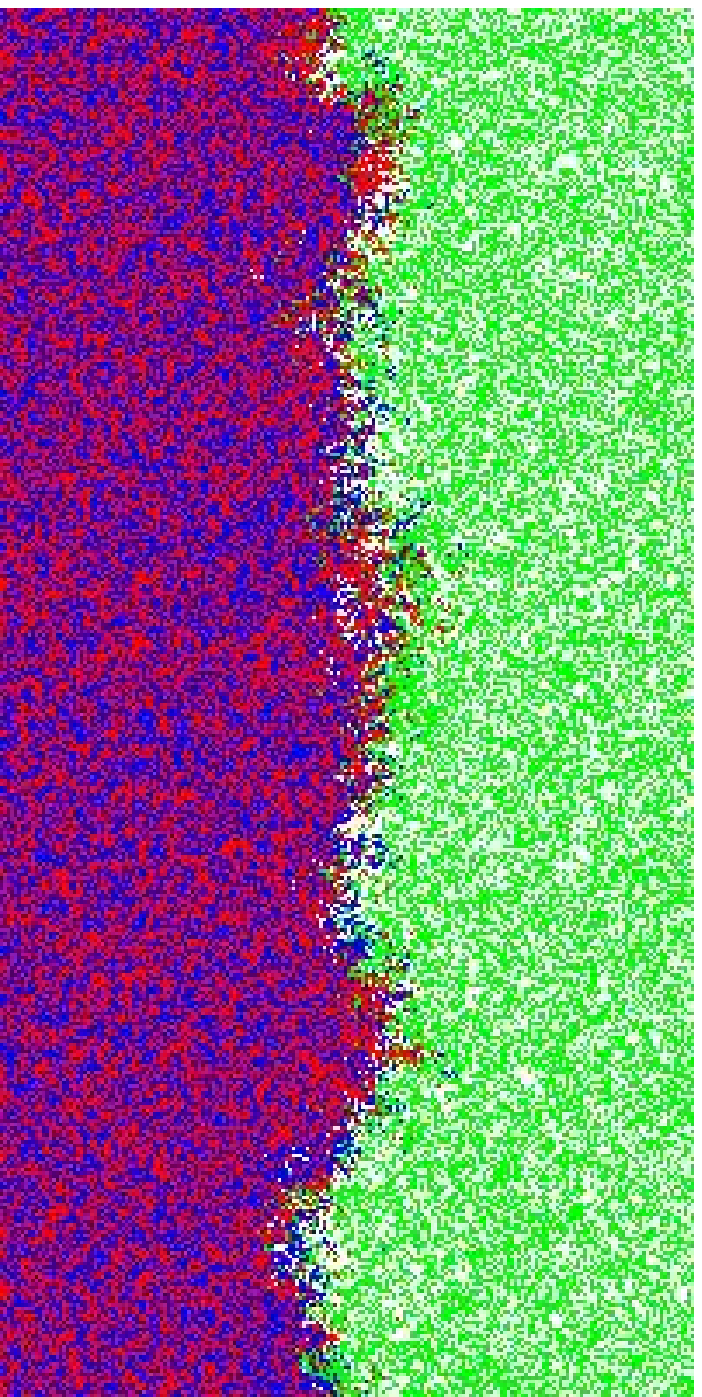}}
\caption{Snapshots of two systems where each half of the system is initially occupied by one alliance.
Left: $k=0.5$, $s=0.5$, and $s_n = 0.5$; right: $k=0.2$, $s=0.8$, and $s_n = 0.8$. The system size is 
$200 \times 400$, the configurations are obtained after 980 MCS.}
\label{fig6}
\end{figure}

In order to elucidate this we prepare our system such that in one half we have only $A$ and $C$ particles, whereas 
the other half is formed by the competing alliance composed of $B$ and $D$ particles. We start with a 
sharp, straight interface, using reflective
boundaries in the direction perpendicular to the interface and periodic boundary conditions parallel to
the interface. We then update the system using the same rules as before and 
monitor the roughening of the interface, see Figure \ref{fig6} for two examples.

Interface fluctuations are usually studied in a quantitative way through the interface width. For large
swapping rates, however, the interface is very diffuse, due to the many exchanges taking place at the
boundaries between the two alliances. In order to determine the local position of the interface, we 
assign the value $+1$ to every $A$ and $C$ particle and the value $-1$ to every $B$ and $D$ particle and
then proceed as in \cite{DeV05} (see also \cite{Mul96}). Considering rectangular systems of $L \times H$
spins, we determine for each row $j$ the value $\ell$ that minimizes the sum
\begin{equation}
v(\ell) = \sum\limits_{i=1}^L \left[ S_{i,j} - p(i-\ell) \right]^2~.
\label{eq}
\end{equation}
Here, $S_{i,j} = \pm 1$, depending on which alliance occupies site $(i,j)$, and $p(u)$ is a step function,
with $p =1$ for $u < 0$ and $p = -1$ for $u > 0$ \cite{DeV05}. The local position of the interface at row $j$,
$\ell(j)$, is then given by the value of $\ell$ that minimizes (\ref{eq}). With this coarse-grained interface
profile at hand, we compute the time-dependent interface width
\begin{equation}
W(t) = \sqrt{ \frac{1}{H} \sum\limits_{j=1}^H \left( \ell(j) -  \overline{\ell} \right)^2}~,
\end{equation}
where $\overline{\ell} = \frac{1}{H} \sum\limits_{j=1}^H \ell(j)$ is the mean position of the fluctuating 
interface at time $t$.

\begin{figure}[h]
\centerline{\epsfxsize=5.25in\ \epsfbox{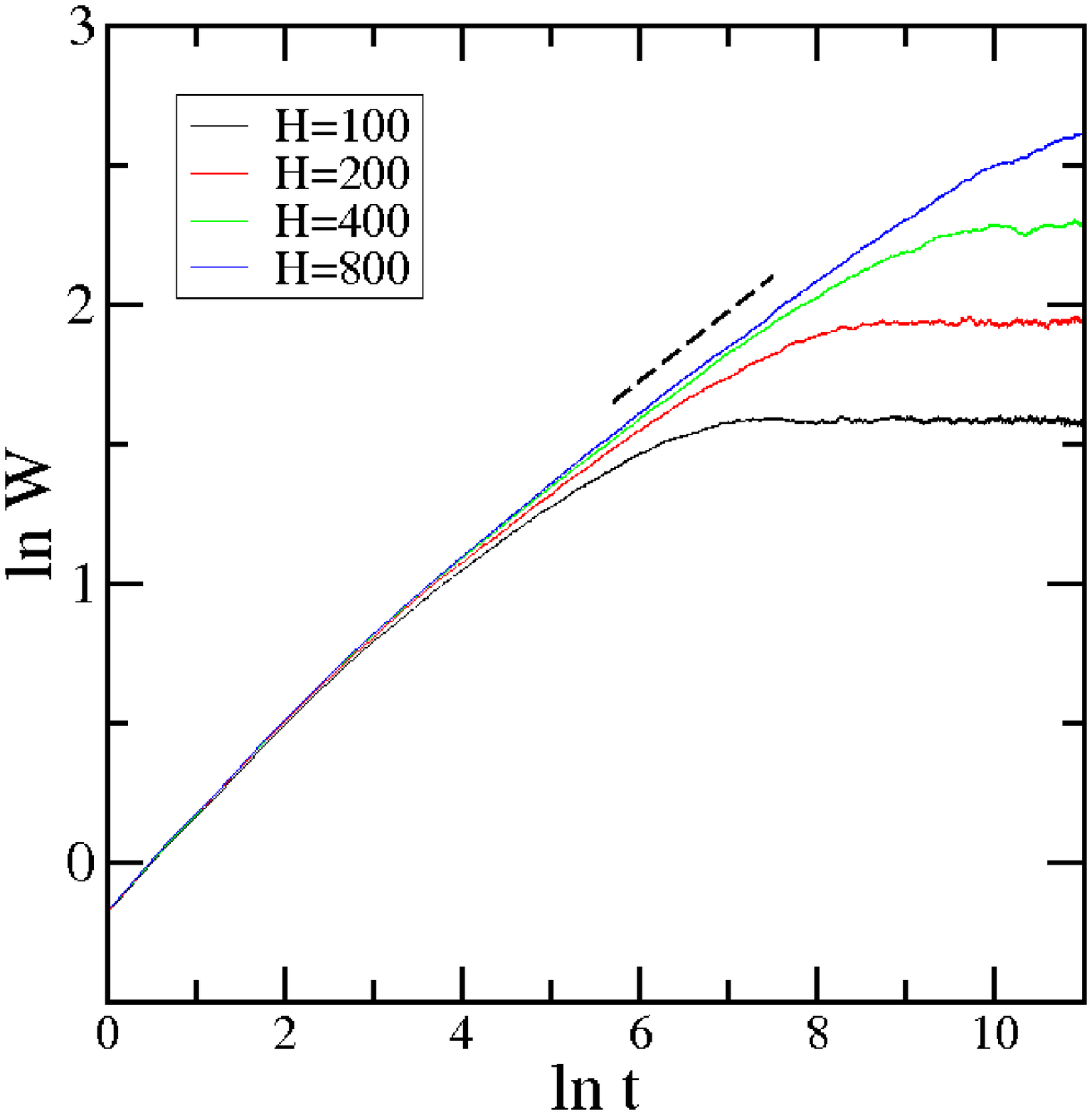}}
\caption{Time evolution of the interface width for the case $k=0.8$, $s=0.2$, and $s_n = 0.2$. Rectangular
systems with $L \times H$ sites are considered, where $L = 200$. The dashed line indicates the expected
power law for correlated fluctuations belonging to the Edwards-Wilkinson universality class. The data result
from averaging over at least 2000 independent realizations of the noise.}
\label{fig7}
\end{figure}

Figure \ref{fig7} shows the time evolution of the interface width for the case $k=0.8$, $s=0.2$, 
and $s_n = 0.2$. We observe three different regimes, as expected for the width of an interface that initially
was given by a straight line: uncorrelated fluctuations prevail at early times, followed by a correlated fluctuation
regime, before the fluctuations saturate at a level that depends on the length $H$ of the interface.
The two last regimes can be summarized by the Family-Vicsek scaling relation \cite{Fam85,Cho09}
\begin{equation}
W(t,H) = H^\alpha f(t/H^{\alpha/\beta})~,
\end{equation}
with the growth exponent $\beta$ and the roughness exponent $\alpha$. It follows from this relation that
the width should grow as $t^\beta$ in the correlated regime and that the value of saturation should scale
as $H^\alpha$ with the interface length. Analyzing our data, we find in the two studied cases that $\beta = 1/4$ (this
is indicated by the dashed line in Figure \ref{fig7}) and $\alpha = 1/2$, in agreement with the values
expected for the Edwards-Wilkinson universality class \cite{Edw82}.

\section{Conclusion}
When four species compete cyclically, the formation of two alliances of mutually neutral partners is observed.
As a result, species extinction, when it takes place, proceeds through a coarsening process where domains formed
by the different alliances expand until one alliance fills the whole system.

In this work we studied the properties of this coarsening process, both in one and two dimensions.
It is known from earlier studies \cite{Fra96a,Fra96b} that in the case of immobile particles the four species 
form domains in one dimension that grow algebraically in time with an exponent 1/3. Mobile particles render the
formation of initial domains more difficult. Once these domains are formed, however, the growth proceeds very fast. If only
predator-prey pairs are allowed to exchange sites, then we obtain an effective power law exponent whose value increases with
the swapping rate. If in addition neutral pairs can swap places, then the domains grow exponentially fast.
In two dimensions the system settles in a steady state characterized by the coexistence of all four species
if no exchanges between neutral partners take place. Allowing for these exchanges, though, yields the formation
of domains formed by partner-pairs, followed by a coarsening process. 
Even so the boundaries between the domains can change their character,
being sharp for small exchange rates, but fuzzy for larger rates, the domain growth exponent takes on the
value 1/2, as expected for curvature-driven coarsening. In addition, after coarse-graining the interface,
we find that the interface fluctuations are characterized by the exponents of the Edwards-Wikinson universality class.

Obviously, the properties of four species that interact in a cyclic way are very different from those encountered
for the case of three species. Three species form in fact a very special situation, as here every species interact
with every other species. In four and more species, however, there are mutually neutral species. As a consequence
one observes for an even number of species the formation of alliances of non-interacting species, thereby enhancing 
the efficiency of the different species to fight off their predators. As this is a very general mechanism,
we expect our results for four species to be representative also for other systems, as long as the
species can arrange themselves in two alliances composed exclusively by partners that are mutually neutral.

\ack
This work was supported by the US National
Science Foundation through grant DMR-0904999.

\end{document}